# Two-Dimensional Multifunctional Materials from Endohedral Fullerenes

Jie Li and Ruqian Wu*

*Department of Physics and Astronomy, University of California, Irvine, California 92697-4575, USA.*

A new multifunctional 2D material is theoretically predicted based on systematic ab-initio calculations and model simulations for the honeycomb lattice of endohedral fullerene $W@C_{28}$ molecules. It has structural bistability, ferroelectricity, multiple magnetic phases, and excellent valley characters and can be easily functionalized by the proximity effect with magnetic isolators such as $MnTiO_3$. Furthermore, we may also manipulate the valley Hall and spin transport properties by selectively switch a few $W@C_{28}$ molecules to the metastable phase. These findings pave a new way in integrate different functions in a single 2D material for technological innovations.

* E-mail: <u>wur@uci.edu</u>.

# I. INTRODUCTION

With diverse new physical properties, two-dimensional (2D) materials have attracted extensive research interest ever since the successful exfoliation and synthesis of graphene [1]. A large variety of 2D materials have been predicted, fabricated and characterized, from insulators (e.g, hexagonal boron nitride [2]), semiconductors (e.g., and silicone [3], germanene [4], phosphorene [5], transition metal dichalcogenides [6], $Cr_2Ge_2Te_6$ [7], $CrI_3$ [8], and stanene [9]), to semimetals (e.g. $TiS_2$ [10], 1T-$MoTe_2$ [11], and $Ta_2Se_3$ [12]). These materials manifest many emergent quantum properties such as high carrier mobility, quantum spin Hall effect (QSHE), quantum anomalous Hall effect (QAHE), valley-polarized anomalous Hall effect (VAHE), and low-dimensional ferromagnetic ordering. However, multifunctional 2D materials with two or more exotic features in a single system have rarely been reported.

Fullerenes, such as $C_{60}$, $C_{72}$, and $C_{80}$, have been extensive studied in the last two decades [13-15]. Research interest in these systems resurges as endohedral fullerenes with additional metal atoms, ions, or clusters embedded in the closed mesh (denoted as M@$C_n$) offer new possibility of producing diverse exotic properties by adjusting the metal cores [16-19]. Small endohedral fullerenes are particularly interesting as they may easily provide environments with low symmetry for metal atoms and hence may develop ferroelectricity, a desired feature for the electric control. Both pristine and endohedral $C_{28}$ molecules have been recently synthesized in experiments [20], they can be hence used as building blocks for the construction of innovative 2D materials [21, 22].

In this paper, we propose a new series of 2D ferroelectric and valleytronic materials: a simple honeycomb lattice of endohedral W@$C_{28}$ molecules. Through systematic ab-initio calculations and model simulations, we found that 2D W@$C_{28}$ honeycomb lattice is an excellent multifunctional 2D material that combines structural bistability, ferroelectricity, multiple magnetic phases, and large valley splitting. Even the ground state of the free-standing W@$C_{28}$ lattice is nonmagnetic, a large valley spin splitting of ~55 meV for VAHE when it is functionalized by a magnetic MnTiO$_3$ substrate. On the other hand, the metastable magnetic phase has sizeable magnetic moment (1 $\mu_B$/molecule) and magnetic anisotropy energy (MAE, 1.09 meV/molecule). As the W atom is ionized in the $C_{28}$ cage, we may switch W@$C_{28}$ molecules between the two phases and manipulate the magnetic state and valley Hall current with a local electric filed, which is ideal for high speed and energy efficient control. The advantage of this material over current valleytronic materials such as transition metal dichalcogenides (TMDs) is obvious as there is no need to change chemical ingredients or apply strong external magnetic field to alter the magnetization and valley splitting. These findings pave a way for developing novel multifunctional 2D vdW materials that are essential for technological innovations.

## II. METHODOLOGY

All the density functional theory (DFT) calculations in this work were carried out with the Vienna ab-initio simulation package (VASP) at the level of the spin-polarized generalized-gradient approximation (GGA) [23]. The interaction between valence electrons and ionic cores was considered within the framework of the projector augmented wave (PAW) method [24, 25]. The energy cutoff for the plane wave basis expansion was set to 500 eV. A Hubbard U = 2.0 eV was added to the 5b orbitals of tungsten (W) for the correlation effect. The vdW correction (DFT-D3) was included in all calculations [26]. All atoms were fully relaxed using the conjugated gradient method for the energy minimization until the force on each atom became smaller than 0.01 eV/Å, and the criterion for total energy convergence was set at $10^{-5}$ eV.

## III. RESULTS AND DISCUSSION

Many endohedral $C_{28}$ ($M@C_{28}$) molecules have been synthesized in carbon vapor [20], which enriched the fullerene chemistry [18, 19] and justifies the material choice in the present work. W is selected as it has large spin orbit coupling (SOC) and possesses magnetic moment in environment with weak interactions. Interestingly, our DFT calculations show that the $W@C_{28}$ molecule has two stable structures (denoted as phase I and II in following discussions). As shown in Fig. 1(a), the W atom displaces by 0.85 Å between these two phases, and the phase I has an energy advantage of 0.131 eV. From the climbing image nudged elastic band (CINEB) calculations, the forward (phase I → II) and backward (phase II → I) transitions have energy barriers of 0.198 eV and 0.067 eV, respectively. DFT calculations further show that the $W@C_{28}$ molecule has a magnetic moment of 2.0 $\mu_B$ in both phases. To determine their magnetic anisotropy energies, we determine the torque $T(\theta)$ as a function of the polar angle $\theta$ between the magnetization and the z axis as shown in Fig. 1(b) [27, 28]. By integrating $T(\theta)$, we obtain the angular dependence of energy, $E(\theta)$. One may see that the highest and lowest energies occur at $\theta=0°$ and $\theta=90°$, indicating that both phases have an in-plane easy axis with the energy differences ($MAE = E(\theta = 90°) - E(\theta = 0°)$) of -0.56 meV and -0.95 meV, respectively. Curves of the projected density of states (PDOS) shown in Fig. 1(c) reveal that the in-plane MAEs mostly stem from the cross-spin SOC interaction between $d_{xy\uparrow}$ and $d_{(x^2-y^2)\downarrow}$ orbitals of tungsten. In addition, it appears that all W d-orbitals hybridize with carbon atoms and the highest occupied molecular orbital (HOMO) of W@C$_{28}$ switches from a basically pure C state in phase I to a W-C intermixed state in phase II. As W@C$_{28}$ molecules are transferable, we expect that they can be used as magnetic superatoms to functionalize other materials such as graphene and topological materials.

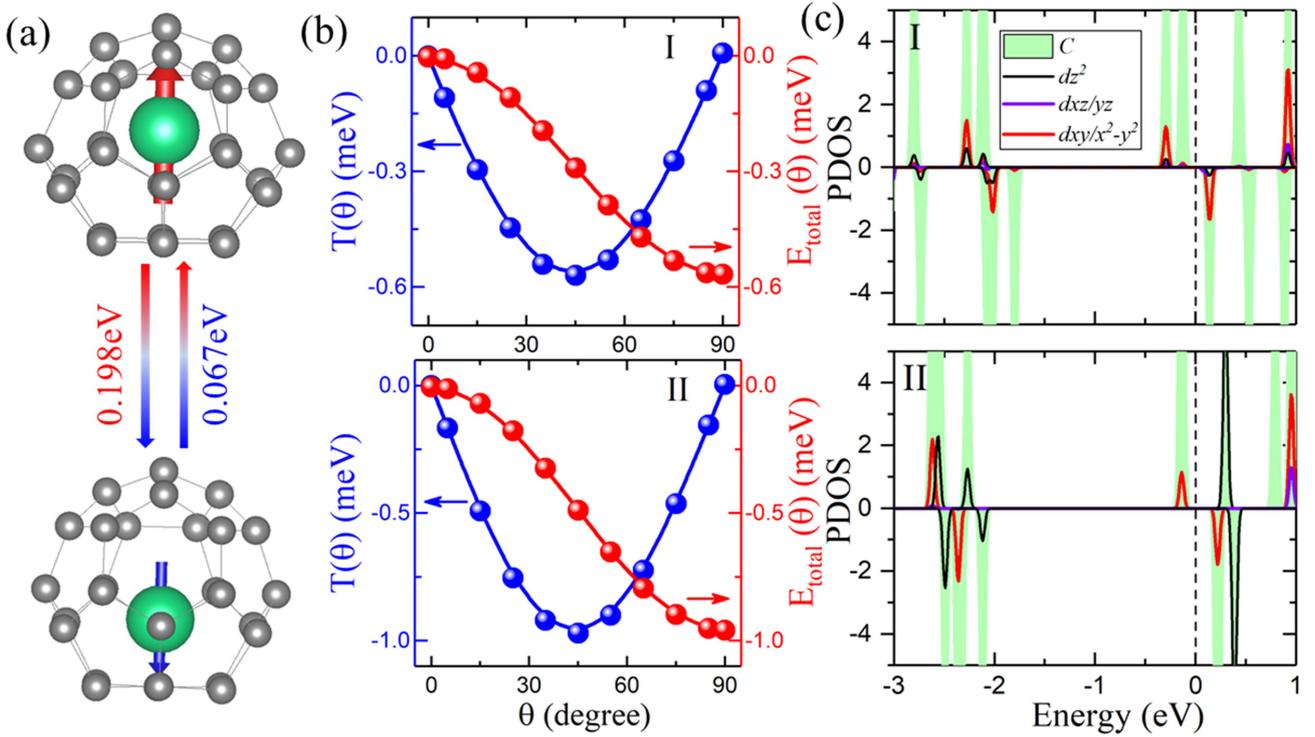

FIG. 1. (a) Schematic structure of W@C$_{28}$ in phases I and II. Two colored arrows represent the direction of the dipoles. (b) Calculated torque vs the angle θ for W@C$_{28}$. (c) The corresponding PDOS of W@C$_{28}$.

In this work, we focus on properties of a W@C$_{28}$ honeycomb lattice as a new functional 2D material. W@C$_{28}$ molecules have the $C_{3v}$ symmetry and it is nature to perceive that they may form either hexagonal or honeycomb structure on appropriate substrates. From our calculations, the honeycomb structure shown in Fig. 2(a) is more stable than the hexagonal structure (see in Fig.

S1 in the Supplemental Material [29]) for a free-standing W@C$_{28}$ monolayer, with a gain of 0.314 eV in binding energy, which is defined as

$$E_b = E_{mol} - E_{2D}/n \qquad (1)$$

Here, $E_{mol}$ and $E_{2D}$ are the total energies of the isolated W@C$_{28}$ and the 2D crystal; and *n* is the number of W@C$_{28}$ molecules per unit cell. Note that the 2D W@C$_{28}$ lattice (2D-W@C$_{28}$) has a binding energy as large as 4.61 eV per molecule, indicating strong interactions among W@C$_{28}$ molecules and decent possibility of synthesizing the 2D-W@C$_{28}$ honeycomb lattice in experiments.

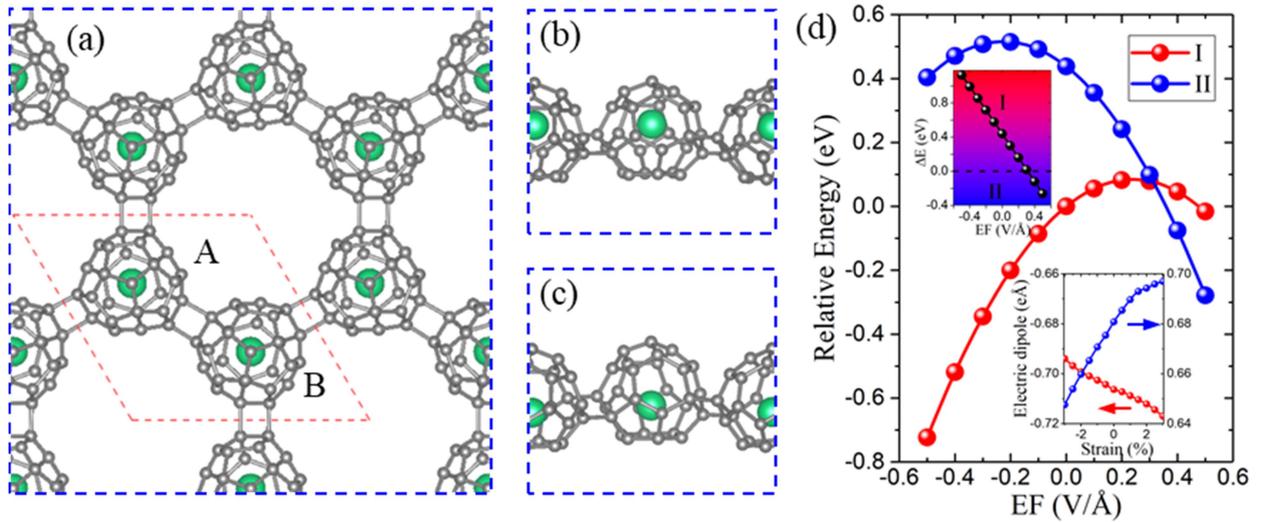

FIG. 2. (a) Top views of 2D-W@C$_{28}$ (the red dashed quadrangle shows the supercell for calculations). (b) (c) Side views of 2D-W@C$_{28}$ in phases I and II, respectively. (d) The relative energy of 2D-W@C$_{28}$ in phases I and II as functions of the external electric field (negative external electric field is along z axis), the inset are the energy differences as functions of the external electric field and the electric dipoles as functions of strain.

The high thermal and dynamic stability of the 2D-W@C$_{28}$ honeycomb lattice is examined by

phonon calculations and ab initio molecular dynamics (AIMD) simulations. The corresponding phonon bands in Fig. S2(a) in the Supplemental Material [29] show the absence of imaginary frequency branch, indicating that the system is dynamically stable. Furthermore, we keep the 2D-W@$C_{28}$ molecules in a 4×4 supercell (928 atoms/supercell) for 10ps (5000 steps) at 300K through AIMD simulations and results show no sign of structure destruction. The total energy fluctuates in a small range without noticeable jump (see Fig. S2(b) in the Supplemental Material [29]). Therefore, we believe that 2D-W@$C_{28}$ honeycomb lattice is thermally stable at least up to room temperature. DFT calculations show that 2D-W@$C_{28}$ also has two stable structures as shown in Fig. 2(c), and the atomic displacement and energy barriers are somewhat enhanced from those for the free molecule, to 0.99 Å and 0.56 eV (0.12 eV for backward switch), respectively (see more details in Fig. S3 in the Supplemental Material [29]). As W donates electrons to the $C_{28}$ cage, there are net electric polarizations in the 2D-W@$C_{28}$ lattice. The calculated electric dipoles for 2D-W@$C_{28}$ are around -0.706 eÅ and 0.683 eÅ per unit cell in phases I and II, respectively. The opposite electric dipole moments indicate that 2D-W@$C_{28}$ is a new ferroelectric material as well. It is striking that the polarization can be switched by a reasonably small external electric field. As shown in Fig. 3(d), a structural phase transition (from I to II) occurs when an external electric field of 0.3 V/Å is applied.

It is known that all ferroelectric materials have piezoelectric properties as well. Here, we further investigate the piezoelectric property of 2D-W@$C_{28}$ by calculating its electric dipole as a function of an in-plane stain. In the inset of Fig. 2(d), one may see that the electric dipoles indeed depend on the strain, particularly for the phase II. Thus, 2D-W@$C_{28}$ as a new ferroelectric material may have a wide range of technological applications, such as non-volatile memories, field effect transistors and sensors.

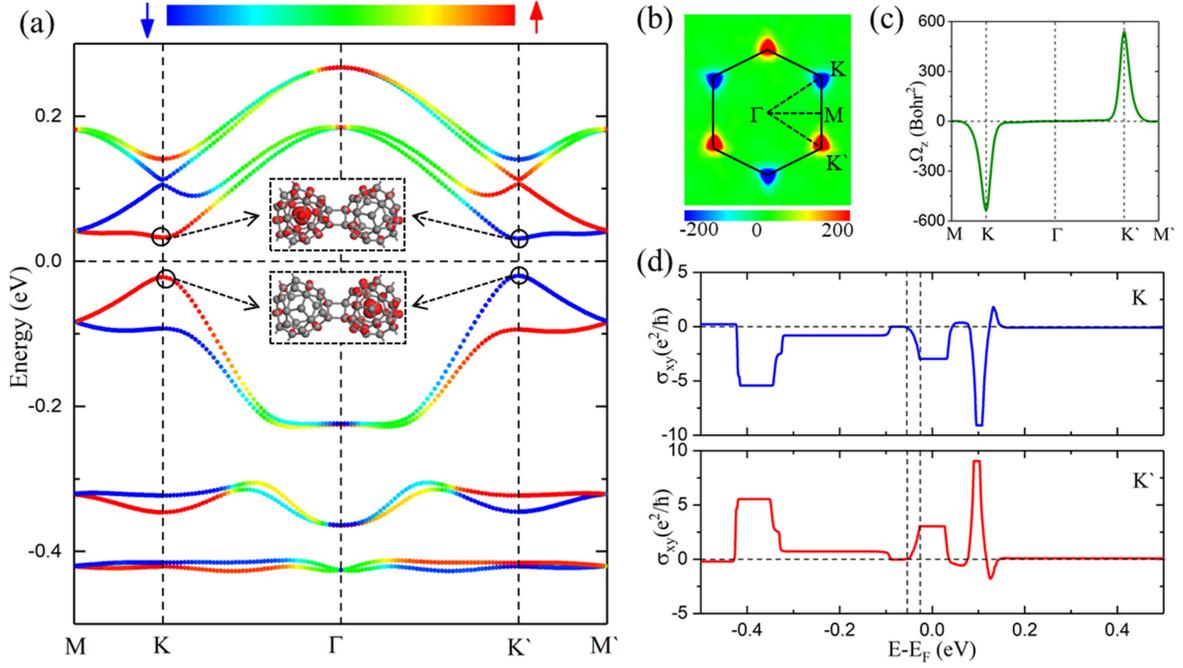

FIG. 3. (a) The band structure of 2D-W@$C_{28}$ with SOC. (Inset are the band decomposed charge density of VBM and CBM at K and K` point, with an isosurface value of 0.001 e/Å$^3$). (b) (c) Calculated Berry curvature of 2D-W@$C_{28}$ over 2D Brillouin zone and along high symmetry lines, respectively. (d) The anomalous Hall conductivity $\sigma_{xy}$ as a function of Fermi level.

In the phase I, the W atom of 2D-W@$C_{28}$ obviously shifts toward the dome and the symmetry is reduced. DFT calculations show that 2D-W@$C_{28}$ is a nonmagnetic semiconductor in phase I, due to the strong intermolecular interactions. Considering the strong SOC of tungsten, we first examine if 2D-W@$C_{28}$ is topologically nontrivial. As shown in Fig. S4 in the Supplemental Material [29], the calculated $Z_2$ is zero so 2D-W@$C_{28}$ is not a topological material by itself. Nonetheless, 2D-W@$C_{28}$ can be an excellent valleytronic material or can become a topological material after being functionalized by others. To examine this possibility, we may analyze its band structure in Fig. 3 and PDOS in Fig. S5 in the Supplemental Material [29]. There is a large band gap of ~55 meV at both K and K` points with opposite spins because of strong SOC of W. The band decomposed charge densities show that holes and electrons residing at K and K` valleys are from the A and B sublattices, respectively. The PDOS curves show that electronic states near to Fermi level contain the $d_{xz/yz}$ orbitals of W and $p_{x/y}$ orbitals of carbon. The map of Berry curvature, $\Omega(k)$, of 2D-W@$C_{28}$ in the 2D Brillouin zone (BZ) and along the high symmetric lines are shown in Fig. 3(b), (c). Obviously, $\Omega(k)$ has the same magnitude but opposite signs around the K and K` points, similar to the feature of pristine TMD monolayers and graphene. Quantitatively, we give the Hall conductivity as a function of the position of the Fermi level by integrating $\Omega(k)$ over vicinities around the K and K` valleys and summing over states below $E_F$, i.e., $\sigma_{xy} = \sum_{n \in occ} \frac{e^2}{\hbar} \int_{K \text{ or } K`} \Omega_n(k) \frac{dk}{(2\pi)^2} = C_{K \text{ or } K`} \frac{e^2}{\hbar}$. Due to the valley degeneracy, $C_K$ and $C_{K`}$ exactly cancel each other in the entire energy range as seen in Fig. 3d, but each of them is nonetheless remarkably close ±3. Obviously, 2D-W@$C_{28}$ has many important features of topological or valleytronic materials and we may attain net valley Hall currents or even QAHE with some modifications.

Among numerous ways of engineering the valley and topological properties, the proximity effect is probably the most intuitive one and has been extensively used in previous studies [30-32]. Here, we use $MnTiO_3$, a potential linear magnetoelectric material, as the substrate to magnetize 2D-W@$C_{28}$. Our DFT calculations show that the bulk $MnTiO_3$ is an antiferromagnetic insulator with a band gap of ~1 eV. A $MnTiO_3$ (001) slab with 12 layers of atoms is used to mimic the substrate, and the surface is terminated by Mn atoms as shown in Fig. 4(a). We use a 2×2 supercell in the lateral plane and fix the in-plane lattice constant to $MnTiO_3$. All atoms except the bottom six MnTiO$_3$ layers are relaxed through DFT calculations. DFT calculations indicate that each Mn atom in the surface layer has a large spin magnetic moment of ~4.5 μ$_B$. This produces a noticeable spin polarization in 2D-W@C$_{28}$, even though the net magnetic moment of each C atom still appear to be small (~0.01μ$_B$).

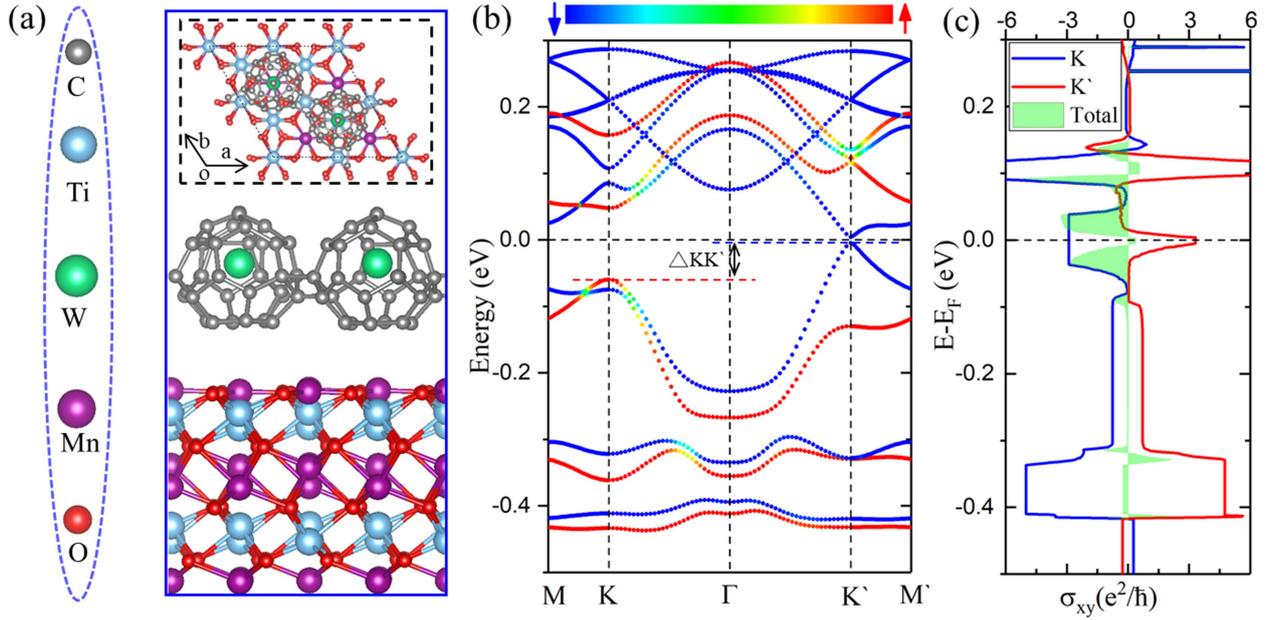

FIG. 4. (a) The side view of structure of 2D-W@C$_{28}$ on MnTiO$_3$ (001) surface. (Inset is the top view). (b) The calculated band structure of 2D-W@C$_{28}$ on MnTiO$_3$ (001) surface. (The valley splitting △KK`= E$_K$-E$_{K`}$, where E$_K$ (E$_{K´}$) represents VBM at K (K´) point). (c) The corresponding anomalous Hall conductivity σ$_{xy}$ as a function of Fermi level.

Let us first focus on the valley splitting. As the valley and spin degrees of freedom are interlocked, the proximity effect of the magnetic MnTiO$_3$ substrate on the two valleys is nonequivalent, as seen in the band structure of 2D-W@C$_{28}$/MnTiO$_3$ in Fig. 4(b). A large valley splitting (△KK`) of ~55 meV can be found as a result of low symmetry of the 2D-W@C$_{28}$ lattice and the inequivalent proximity effect on two spin channels. The K and K' valleys have band gaps of ~100 meV and ~8.8 meV, respectively (as GGA functionals mostly underestimate band gaps, we expect that the actual band gap should be somewhat larger than these numbers). Because the degeneracy between two valleys is broken, we may selectively use carriers from different valleys, e.g., by shifting the Fermi level to below the valence band maximum (VBM) at the K or K` points with a bias. To shed some light on the large valley splitting due to the interplay of the proximity effect, broken symmetry and strong SOC, we construct a *k. p* model for 2D-W@$C_{28}$ (see more details in the Supplemental Material [29]). By fitting the DFT results, the corresponding parameters are extracted as shown in Table S1 in the Supplemental Material [29]. The strong SOC parameters ($\lambda_V$, $\lambda_C$) in both valence band and conduction band are mainly from contributions of W atoms. Negative $\lambda_C$ changes the spin orders between the valence band and conduction band [33]. The large effective exchange fields for valence bands show the robust proximity effect from the substrate, which induces a sizable valley splitting.

A net orbital magnetic moment ($m_v$) associated with the valley index [34] is plotted as a function of the position of the Fermi level in Figure S6 in the Supplemental Material [29]. Obviously, the valley-polarized anomalous Hall effect is analogous to the QAHE and falls into the same category as the Berry-phase related topological transport phenomena. The anomalous Hall conductivity as a function of the position of the Fermi level is shown in Fig. 4(c). Note that the net Chern number takes a value of -3 as we sweep the Fermi level up and down by merely 0.02 eV so the QAHE can be easily realized in this system.

DFT calculations show that 2D-W@$C_{28}$ in phase II is metallic and has a magnetic moment of ~2 $\mu_B$ per unit cell (see in the band structure of 2D-W@$C_{28}$ in phase II in Fig. S7 in the Supplemental Material [29]) and a large MAE of ~2.18 meV per unit cell as shown in Fig. S8(a) in the Supplemental Material [29]. By using the renormalized spin-wave theory (RSWT) [35, 36], we calculate its Curie temperature (the renormalized magnetization M(T)/M(0) as a function of T is shown in Fig. S8(b) in the Supplemental Material [29], and $T_c$ is determined by the location where the renormalized magnetization drops zero). One may see that 2D-W@$C_{28}$ in phase II has exceedingly high Curie temperatures of ~167K, a new van der Waals type metallic 2D magnetic material. Furthermore, we may use W@$C_{28}$ molecules in phase II as magnetic dopants for the control of transport properties of the phase-I W@$C_{28}$ lattice. To this end, we propose a conceptual device shown in Fig. S9 in the Supplemental Material [29] to produce the VAHE. Using silicon as the substrate as well as the channel for applying the bias and a hexagonal BN (hex-BN) overlayer to offer a protective environment, a few W@$C_{28}$ molecules in the Hall bar can be selectively switched to the phase II by using local electric field. When the distance between adjacent magnetic molecules is shorter than the spin mean free path, they are expected to produce valley anomalous Hall effect with the same mechanism as discussed above for 2D-W@$C_{28}$/MnTiO$_3$. This idea can be proven to some extent by the band structure of 2D-W@$C_{28}$ in Fig. S10 in the Supplemental Material [29] with 25% molecules in the phase II, in which the degeneracy between the two valleys is clearly lifted. This is very different from other valleytronic materials such as TMDs for which large external magnetic field or optical excitation is needed to break the valley degeneracy. Obviously, the distribution and density of "dopants" in this conceptual device can be easily reprogramed to change the transport properties.

## IV. CONCLUSION

In summary, we proposed a new 2D ferroelectric and valleytronic material based on endohedral fullerenes $C_{28}$ and further explored its structural and electronic properties through systematic ab-initio calculations and model simulations. The results show that 2D-W@$C_{28}$ has structural bistability, ferroelectricity, multiple magnetic phases, and excellent valley characters. In the ground state nonmagnetic phase, a large valley spin splitting of ~55 meV can be induced by the proximity effect from the $MnTiO_3$ substrate, which offers the possibility for developing a new series of valleytronic materials. Furthermore, we proposed a new strategy to use local electric field for the manipulation of valley Hall current, based on the exotic ferroelectric property and multiple magnetic phases of W@$C_{28}$. Note that W@$C_{28}$ molecules are chemically active and may self-assemble to a honeycomb lattice due to the appreciable energy preference over the close packed structure. For the same reason, a protective layer such as hBN might be needed for the use in ambient condition. We hope that our studies may inspire more experimental explorations so this novel 2D multifunctional material or related systems can be developed for diverse applications.

## ACKNOWLEDGMENTS

Work was supported by US DOE, Basic Energy Science (Grant No. DE-FG02-05ER46237). Calculations were performed on parallel computers at NERSC.